\documentclass[sigconf, nonacm]{acmart}

\usepackage{amsmath}
\usepackage{booktabs}
\usepackage{multirow}
\usepackage{algorithm}
\usepackage{algorithmic}
\usepackage{xcolor}
\usepackage{tcolorbox}
\tcbuselibrary{listings,skins,breakable}
\usepackage{listings}
\usepackage{xcolor,colortbl}
\usepackage{ulem}
\usepackage{graphicx}
\usepackage{newunicodechar}

\AtBeginDocument{%
  }
    
\setcopyright{none}  % 移除版權聲明
\renewcommand\footnotetextcopyrightpermission[1]{}  % 移除第一欄的版權註腳
\begin{document}
\title{CGPT: Cluster-Guided Partial Tables with LLM-Generated Supervision for Table Retrieval}

\author{Tsung-Hsiang Chou}
\affiliation{%
  \institution{National Chung Hsing University}
  \institution{Smart Sustainable New Agriculture Research Center (SMARTer)}
  \city{Taichung}
  \country{Taiwan}}
\email{yumeow0122@smail.nchu.edu.tw}

\author{Chen-Jui Yu}
\affiliation{%
  \institution{National Chung Hsing University}
  \institution{Smart Sustainable New Agriculture Research Center (SMARTer)}
  \city{Taichung}
  \country{Taiwan}}
\email{rui0828@smail.nchu.edu.tw}

\author{Shui-Hsiang Hsu}
\affiliation{%
  \institution{National Chung Hsing University}
  \institution{Smart Sustainable New Agriculture Research Center (SMARTer)}
  \city{Taichung}
  \country{Taiwan}}
\email{g113056055@smail.nchu.edu.tw}

\author{Yao-Chung Fan}
\affiliation{%
  \institution{National Chung Hsing University}
  \institution{Smart Sustainable New Agriculture Research Center (SMARTer)}
  \city{Taichung}
  \country{Taiwan}}
\email{yfan@nchu.edu.tw}

\begin{abstract}
General-purpose embedding models have demonstrated strong performance in text retrieval but remain suboptimal for table retrieval, where highly structured content leads to semantic compression and query–table mismatch. Recent LLM-based retrieval augmentation methods mitigate this issue by generating synthetic queries, yet they often rely on heuristic partial-table selection and seldom leverage these synthetic queries as supervision to improve the embedding model. We introduce \textsc{CGPT}, a training framework that enhances table retrieval through \textit{LLM-generated supervision}. \textsc{CGPT} constructs semantically diverse partial tables by clustering table instances using K-means and sampling across clusters to broaden semantic coverage. An LLM then generates synthetic queries for these partial tables, which are used in hard-negative contrastive fine-tuning to refine the embedding model. Experiments across four public benchmarks (MimoTable, OTTQA, FetaQA, and E2E-WTQ) show that \textsc{CGPT} consistently outperforms retrieval baselines, including QGpT, with an average R@1 improvement of 16.54\%. In a unified multi-domain corpus setting, \textsc{CGPT} further demonstrates strong cross-domain generalization and remains effective even when using smaller LLMs for synthetic query generation. These results indicate that semantically guided partial-table construction, combined with contrastive training from LLM-generated supervision, provides an effective and scalable paradigm for large-scale table retrieval. Our code is available at \url{https://github.com/yumeow0122/CGPT}.

\end{abstract}




\maketitle
\section{Introduction}

Tables are a central medium for storing and communicating structured information across domains such as finance, science, logistics, and public knowledge bases. With the rapid growth of large web-scale table corpora, effective table retrieval has become increasingly important for downstream tasks, e.g., table question answering. However, general-purpose embedding models trained on textual corpora exhibit strong generalization in text retrieval, yet face substantial challenges when extended to table retrieval. Encoding an entire table into a single vector thus induces semantic compression, resulting in degraded retrieval performance. Enhancing representation learning~\cite{herzig2021open,pan2021cltr,liang2025improving} for table retrieval is therefore a central objective.

Beyond task-specific table embedding fine-tuning~\cite{herzig2021open,pan2021cltr}, recent work has explored the use of large language models (LLMs) to provide additional supervision for table retrieval without modifying the embedding model directly. A representative approach in this direction is QGpT~\cite{liang2025improving}, which selects the first $k$ rows of a table to form a partial table and prompts an LLM to generate \textit{synthetic queries} (SQ). These synthetic queries serve as enhanced table representations and can improve retrieval accuracy.

While effective, QGpT has two key limitations:  
(1) selecting only the first rows assumes they are representative, which does not hold when relevant information appears in later rows; and  
(2) the generated synthetic queries are seldom used as \textit{training supervision} to improve the embedding model itself, leaving potential performance gains unrealized.

\begin{figure*}
    \centering
    \includegraphics[width=\textwidth]{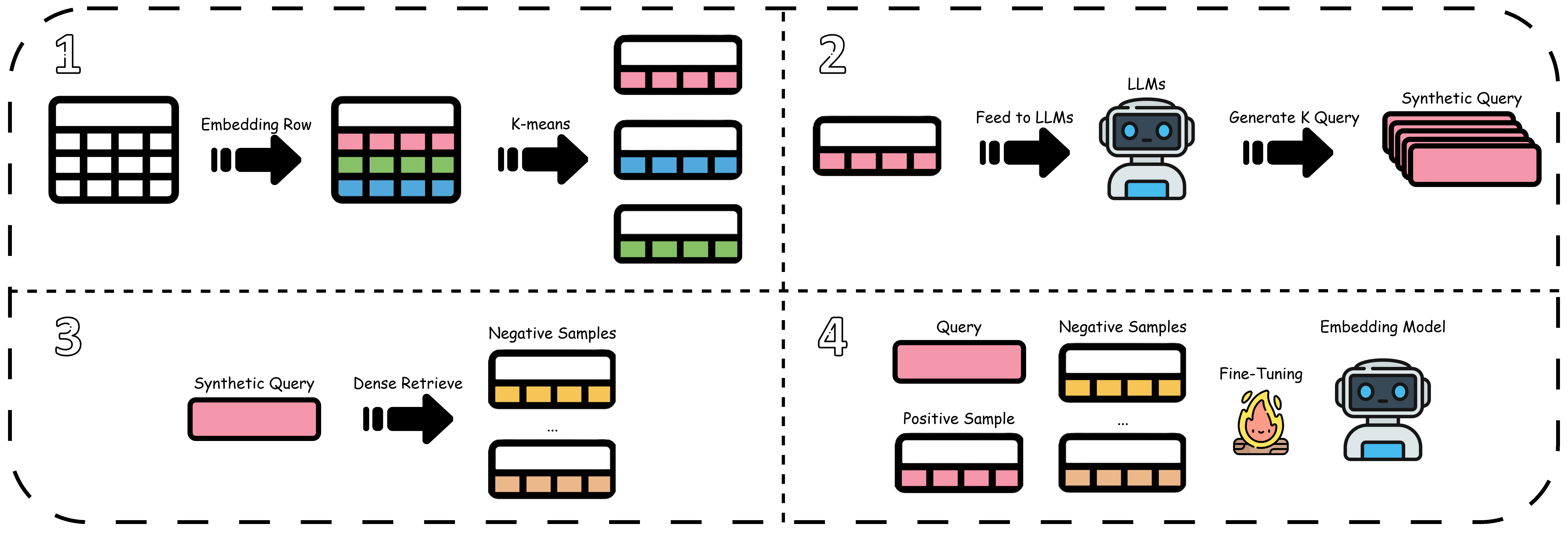}
    \caption{Overall workflow of \textsc{CGPT}. The framework consists of four main stages: (1) clustering-based partial table generation; (2) synthetic query generation; (3) hard negative sampling; and (4) model fine-tuning.}
    \label{fig:workflow}
\end{figure*}

To address these challenges, we propose \textsc{CGPT}, a framework that enhances table retrieval through \textbf{LLM-generated supervision}. \textsc{CGPT} applies K-means clustering to group table rows into semantically coherent subsets and samples from each cluster to construct partial tables with broader coverage of table attributes and instances. Synthetic queries generated by an LLM are then leveraged in hard-negative contrastive fine-tuning, enabling the embedding model to better retrieve tables even when relevant information is sparsely located across rows.

We evaluate \textsc{CGPT} on four public benchmarks—MimoTable, OTTQA, FetaQA, and E2E-WTQ—and observe consistent improvements over retrieval augmentation baselines, with a mean R@1 gain of 16.54\%. Additional experiments on a unified multi-domain corpus confirm cross-domain generalization, and experiments with smaller LLMs show that \textsc{CGPT} remains cost-efficient without sacrificing retrieval effectiveness.

\section{Related Work}
\subsection{Table Retrieval}
Table retrieval has emerged as an important problem in natural language processing. TAPAS~\cite{herzig2020tapas} extends the BERT architecture with table-aware positional encodings (e.g., row and column indices) and pre-trains on millions of Wikipedia tables, establishing a strong foundation for table QA. Dense Table Retrieval (DTR)~\cite{herzig2021open} adopts dense vector retrieval by encoding queries and tables separately with two TAPAS encoders and computing similarity via inner products. DTR further incorporates the Inverse Cloze Task~\cite{lee2019latent} and hard negative sampling to improve retrieval quality.

A common limitation of these approaches is the use of a single vector to represent an entire table. For large tables, this design can cause excessive semantic compression.

\subsection{Representation Augmentation}
Recent work explores synthetic query generation to strengthen table representations. QGpT~\cite{liang2025improving} adopts a partial-table strategy by selecting the first 10 rows and generating synthetic queries using a large language model; the partial table and generated queries jointly define the table representation. This reduces input length and yields more focused supervision, improving zero-shot performance of general-purpose embedding models. GenSearch~\cite{khurana2025table} instead generates pseudo tables conditioned on queries, further illustrating the promise of generative retrieval.

Building on these ideas, we aim to construct partial tables that more comprehensively cover the semantic space of the original table. Our method, \textsc{CGPT}, clusters table instances via K-means and samples across clusters to form semantically diverse partial tables with improved coverage.

\section{Method}

As illustrated in Figure~\ref{fig:workflow}, \textsc{CGPT} follows a four-stage training pipeline: clustering–based partial table generation, synthetic query generation, hard negative sampling, and contrastive fine-tuning. Given a table $T = \{H, I_1, \dots, I_m\}$ with header $H$ and $m$ instances, we construct training data and optimize the embedding model through the steps below.

\subsection{Clustering-based Partial Table Generation}
To capture a broader portion of the table’s semantic space, we generate multiple representative partial tables using K-means, called KPT (K-means Partial Table). This process includes instance embedding, clustering, and partial-table formation.

\textbf{Instance Embedding} 
For each instance $I_i$ in table $T$, we encode it using a pretrained embedding model, producing a set of instance embedding vectors  
$\mathcal{e}=\{\mathbf{e}_1, \mathbf{e}_2, ..., \mathbf{e}_m\}$.

\textbf{K-means Clustering}
To ensure that the constructed partial tables capture the table’s full semantic space, we apply K-means clustering over $\mathcal{e}$.  
The number of clusters $k$ is determined adaptively based on the table size:

\begin{equation}
k = \min\left(\left\lceil \frac{m}{r} \right\rceil, k_{\max}\right)
\end{equation}

\textbf{Forming K-means Partial Table}
From each cluster $C_j$, we randomly sample $s$ instances to construct the $j$-th partial table:

\begin{equation}
KPT_j = \{H\} \cup \{I_i \mid I_i \in \text{Sample}(C_j, s)\}
\end{equation}

This procedure yields $k$ partial tables per original table, with each KPT reflecting the semantics of a specific cluster.

\subsection{Synthetic Query Generation}
For each partial table $KPT_j$, we prompt a LLM to generate $n_q$ synthetic queries. The generation process is model-agnostic and only requires the ability to produce natural-language queries conditioned on the content of the partial table. For reproducibility, the exact prompt template is provided in Appendix~\ref{sec:prompts}.

\subsection{Hard Negative Sampling and Fine-tuning}
To strengthen the model’s ability to distinguish semantically similar but incorrect tables, we integrate hard negative sampling with contrastive fine-tuning.

\subsubsection{Hard Negative Sampling}
For each synthetic query $q$, we compute cosine similarity between $q$ and all KPTs from other tables using a pretrained embedding model, selecting the top-$h$ most similar but incorrect ones as hard negatives to form $\mathcal{HN}$. These negatives, which are easily confused with the true positive, provide strong supervision for improving the model's discrimination.

\subsubsection{Fine-tuning}
Given a synthetic query $q$, its corresponding positive KPT $p^+$, and the hard negative set $\mathcal{HN}$, we construct training triples $(q, p^+, \mathcal{HN})$ and optimize the embedding model using the InfoNCE objective~\cite{oord2018representation}:

\begin{equation}
\mathcal{L} = - \log \frac{\exp(\text{sim}(q, p^+) / \tau)}{\exp(\text{sim}(q, p^+) / \tau) + \sum_{p^- \in \mathcal{HN}} \exp(\text{sim}(q, p^-) / \tau)}
\end{equation}

The symbols are defined as follows:
\begin{itemize}
    \item $q$ denotes the synthetic query.
    \item $p^+$ is the corresponding positive KPT.
    \item $p^- \in \mathcal{HN}$ are the selected hard-negative KPTs.
    \item $\text{sim}(\cdot,\cdot)$ denotes cosine similarity.
    \item $\tau$ is a tunable temperature parameter.
\end{itemize}

\section{Experiment}

We evaluate \textsc{CGPT} on four benchmarks: MimoTable (containing both Chinese and English tables), OTTQA, FetaQA, and E2E-WTQ. 

\subsection{Experiment Setting}
During partial table generation, three parameters control semantic coverage: the cluster granularity parameter $r$ is set to 10, the maximum number of clusters is $k_{\max}=5$, and $s=5$ instances are sampled from each cluster to build the final KPTs.

For synthetic query generation, we use \texttt{Llama-3.1-8B-Instruct} to produce queries, generating $n_q=5$ queries per partial table. The generation parameters include a temperature of $0.4$ and a maximum output length of 1024 tokens. For hard negative sampling, we select $h=8$ hard negatives per query.

For model fine-tuning, we adopt \texttt{BAAI/bge-m3} as the base embedding model. Training uses a learning rate of $1\text{e-}5$, two epochs, a temperature $\tau = 0.01$, and gradient accumulation with 32 steps. All experiments are conducted on a single NVIDIA A6000 GPU with 48GB memory.

\subsection{Main Result}

\begin{table*}
\centering
\small
\caption{Comparison of retrieval performance across four benchmark datasets.}
\label{tab:main_results}
\begin{tabular}{cccccccccccccccc}
    \toprule
    \multirow{2}{*}{\parbox{2cm}{\centering Method}}
    & \multicolumn{3}{c}{MimoTable (CH)} & \multicolumn{3}{c}{MimoTable (EN)} & \multicolumn{3}{c}{OTTQA} & \multicolumn{3}{c}{FetaQA} & \multicolumn{3}{c}{E2E-WTQ} \\
    \cmidrule(lr){2-4} \cmidrule(lr){5-7} \cmidrule(lr){8-10} \cmidrule(lr){11-13} \cmidrule(lr){14-16}
    & R@1 & R@5 & R@10 & R@1 & R@5 & R@10 & R@1 & R@5 & R@10 & R@1 & R@5 & R@10 & R@1 & R@5 & R@10 \\
    \midrule
    QGpT & 50.6 & 75.0 & 81.45 & 50.66 & 72.35 & 80.8 & 51.45 & 78.14 & 86.68 & 33.95 & 50.87 & 57.86 & 41.49 & 65.98 & 72.61 \\
    \midrule
    CGPT w/o FT & 52.74 & 76.64 & 83.23 & 57.14 & 79.46 & 85.17 & 84.24 & \textbf{97.24} & 98.46 & 34.85 & \textbf{52.82} & 60.91 & 69.29 & 91.7 & \textbf{96.68}\\
    % CGPT w/o Clustering & 48.91 & 70.41 & 79.28 & 53.45 & 76.66 & 81.72 & 84.73 & 97.19 & \textbf{98.59} & 26.75 & 45.43 & 54.81 & 58.92 & 87.13 & 91.7 \\
    CGPT w/o HNS & 55.51 & \textbf{80.57} & \textbf{85.66} & 57.84 & 80.07 & 85.28 & 84.69 & 97.02 & \textbf{98.55} & 34.85 & 52.32 & \textbf{61.11} & 68.05 & \textbf{92.12} & 96.27 \\
    CGPT & \textbf{56.8} & 79.81 & 84.2 & \textbf{60.13} & \textbf{81.49} & \textbf{85.65} & \textbf{86.86} & 97.07 & 98.33 & \textbf{34.9} & 51.52 & 59.11 & \textbf{72.2} & 91.7 & 95.44 \\
    \bottomrule
\end{tabular}
\end{table*}

Table~\ref{tab:main_results} presents the retrieval performance of \textsc{CGPT} and baseline systems across four benchmarks. We conduct a systematic ablation study to isolate the contributions of K-means clustering and hard negative sampling. The comparison includes QGpT (using its original configuration and data), \textsc{CGPT} w/o FT (no fine-tuning), and \textsc{CGPT} w/o HNS (replacing hard negatives with random negatives).

K-means clustering provides a clear advantage even without fine-tuning, \textsc{CGPT} w/o FT achieves R@1 scores of 52.74\% and 57.14\% on the two compared subsets, improving over QGpT by 2.14 and 6.48 points, respectively. This demonstrates that semantically guided partial-table construction alone yields stronger table representations, underscoring the value of cluster-aware table generation.

Regarding sampling strategies, \textsc{CGPT} w/o HNS attains the best R@5 and R@10 scores on MimoTable (CH), showing that the semantic diversity introduced by K-means is sufficient to support strong top-5/10 retrieval even with simplified negative sampling.

For precision-focused evaluation, the full \textsc{CGPT} method shows the strongest R@1 performance. On MimoTable (EN), \textsc{CGPT} reaches 60.13\% R@1, surpassing QGpT by 9.47 points. Similar improvements are observed on more complex datasets such as OTTQA and E2E-WTQ. Although hard negative sampling may slightly trade off top-5/10 performance, it improves top-1 score by providing more informative contrastive signals. These results indicate that while clustering-based partial table construction enhances semantic coverage, combining it with hard negative sampling is essential for maximizing precision in table retrieval tasks.

\subsection{Evaluation on the QGpT Dataset}

Table~\ref{tab:qgpt_original} presents the performance of the \textsc{CGPT}-trained model on the original QGpT dataset, allowing us to assess cross-strategy transfer given the differing partial-table construction methods.

\textsc{CGPT} shows clear improvements: MimoTable (EN) R@1 increases from 50.66\% to 59.28\%, and MimoTable (CH) from 50.6\% to 53.54\%. These results indicate that the semantic representations learned by \textsc{CGPT} transfer effectively across construction schemes, and that its clustering-based modeling offers robustness to semantically dispersed tables, supporting broad applicability in retrieval settings.

\begin{table}[H]
\centering
\small
\caption{Retrieval performance of \textsc{CGPT} on the original QGpT dataset.}
\label{tab:qgpt_original}
\resizebox{\columnwidth}{!}{%
\begin{tabular}{lcccccc}
    \toprule
    \multirow{2}{*}[-0.7ex]{\centering Method} & \multicolumn{3}{c}{MimoTable (CH)} & \multicolumn{3}{c}{MimoTable (EN)} \\
    \cmidrule(lr){2-4} \cmidrule(lr){5-7}
    & R@1 & R@5 & R@10 & R@1 & R@5 & R@10 \\
    \midrule
    QGpT   & 50.6  & 75.0  & 81.45 & 50.66 & 72.35 & 80.8  \\
    QGpT w/ CGPT & \textbf{53.54} & \textbf{77.27} & \textbf{82.75} & \textbf{59.28} & \textbf{80.29} & \textbf{83.98} \\
    \bottomrule
    \end{tabular}%
}
\end{table}

\begin{table*}
\centering
\small
\caption{Comparison of retrieval performance on the mixed dataset.}
\label{tab:full_results}
\begin{tabular}{cccccccccccc}
    \toprule
    \multirow{2}{*}{\parbox{1.3cm}{\centering Model}} & \multirow{2}{*}{\parbox{1.3cm}{\centering Strategy}}
    & \multicolumn{2}{c}{MimoTable (CH)} & \multicolumn{2}{c}{MimoTable (EN)} & \multicolumn{2}{c}{OTTQA} & \multicolumn{2}{c}{FetaQA} & \multicolumn{2}{c}{E2E-WTQ} \\
    \cmidrule(lr){3-4} \cmidrule(lr){5-6} \cmidrule(lr){7-8} \cmidrule(lr){9-10} \cmidrule(lr){11-12}
    & & R@1 & R@5 & R@1 & R@5 & R@1 & R@5 & R@1 & R@5 & R@1 & R@5 \\
    \midrule
    \multirow{2}{*}{\parbox{1.3cm}{\centering BGE-m3}}
    & QGpT & 38.54 & 62.67 & 39.46 & 61.64 & 82.02 & 95.58 & 30.3 & 46.38 & 35.27 & 61 \\
    & KPT & 43.14 & 65.0 & 40.99 & 63.6 & \textbf{84.64} & \textbf{96.21} & 29.16 & 45.98 & 53.94 & \textbf{74.27} \\
    \midrule
    \multirow{2}{*}{\parbox{1.3cm}{\centering CGPT}}
    & QGpT & 55.03 & 75.57 & \textbf{57.79} & 77.78 & 78.36 & 92.59 & 32.0 & \textbf{47.58} & 36.93 & 58.09 \\
    & KPT & \textbf{57.2} & \textbf{78.18} & 57.76 & \textbf{77.95} & 80.22 & 94.08 & \textbf{32.85} & 47.68 & \textbf{56.85} & 73.44 \\
    \bottomrule
\end{tabular}
\end{table*}

\subsection{Experiment on Unified Multi-Domain Dataset}
To assess the cross-domain generalization capability of \textsc{CGPT}, we merge all datasets into a single unified multi-domain corpus and conduct large-scale retrieval evaluation. The QGpT baseline is similarly tested on the merged corpus. Table~\ref{tab:full_results} reports the performance of QGpT, KPT, the BGE-M3 model, and the \textsc{CGPT}-trained model.

On the unfine-tuned BGE-M3 model, KPT yields notable gains. For example, on MimoTable (CH), it achieves 43.14\% R@1, outperforming QGpT by 4.6 points. On E2E-WTQ, KPT reaches 53.94\% R@1, surpassing QGpT’s 35.27\% by 18.67 points. These results indicate that KPT effectively preserves key table semantics and enhances retrieval accuracy even under multi-domain conditions.

With \textsc{CGPT}, the improvements are further amplified. On MimoTable (EN), \textsc{CGPT} boosts R@1 to 57.79\%, an 18.33-point increase over the BGE-M3 baseline; on MimoTable (CH), it reaches 55.03\%, improving by 16.49 points. When combined with KPT, \textsc{CGPT} achieves 57.20\% R@1 and 78.18\% R@5 on MimoTable (CH). These results demonstrate that \textsc{CGPT} maintains strong and stable performance across diverse domains and languages, exhibiting robust cross-domain generalization suitable for large-scale table retrieval applications.

\subsection{Comparison Across Different LLMs}

Table~\ref{tab:different_llms} reports the performance of different LLMs. We compare three LLMs of varying scales:  \texttt{Llama-3.1-8B-Instruct}, \texttt{GPT-OSS-20B}, and \texttt{Qwen3-4B} to assess their impact on \textsc{CGPT}’s retrieval performance on MimoTable (EN). Despite differences in model size and architecture, the results remain highly consistent, with R@1 varying by only 0.6 percentage points. This indicates that \textsc{CGPT} is robust to the choice of language model, allowing practitioners to adopt smaller, more cost-efficient models without sacrificing retrieval effectiveness.

\begin{table}
\caption{Comparison Across Different LLMs}
  \centering
  \small
  \resizebox{\columnwidth}{!}{%
  \begin{tabular}{lcccccc}
    \toprule
    \multirow{2}{*}{\parbox{1.5cm}{\centering Method}} & \multicolumn{3}{c}{MimoTable (CH)} & \multicolumn{3}{c}{MimoTable (EN)} \\
    \cmidrule(lr){2-4} \cmidrule(lr){5-7}
    & R@1 & R@5 & R@10 & R@1 & R@5 & R@10 \\
    \midrule
    CGPT (Llama-3.1-8B) & 56.8 & \textbf{79.81} & 84.2 & \textbf{60.13} & \textbf{81.49} & 85.65 \\
    w/ GPT-OSS-20B & 55.3 & 77.48 & 81.94 & 59.53 & 80.58 & \textbf{85.92} \\
    w/ Qwen3-4B & \textbf{58.66} & 78.93 & \textbf{84.29} & 60.1 & 80.75 & 84.8 \\
    \bottomrule
  \end{tabular}%
  }
  \label{tab:different_llms}
\end{table}

\subsection{Ablation Study: Instance Sampling}
\begin{table}
  \centering
  \small
  \caption{Comparison of retrieval performance for different instance sampling strategies}
  \resizebox{\columnwidth}{!}{%
  \begin{tabular}{lcccccc}
    \toprule
    \multirow{2}{*}{\parbox{1.5cm}{\centering Method}} & \multicolumn{3}{c}{MimoTable (CH)} & \multicolumn{3}{c}{MimoTable (EN)} \\
    \cmidrule(lr){2-4} \cmidrule(lr){5-7}
    & R@1 & R@5 & R@10 & R@1 & R@5 & R@10 \\
    \midrule
    CGPT & 56.8 & 79.81  & 84.2 & \textbf{60.13} & \textbf{81.49} & \textbf{85.65}  \\
    w/ CB Selection & 51.62 & 70.55 & 76.58 & 57.51 & 79.65 & 83.78 \\
    w/ S Selection & \textbf{58.13} & \textbf{80.11} & \textbf{84.99} & 57.03 & 80.43 & 85.64 \\
    \bottomrule
  \end{tabular}%
  }
  \label{tab:ablation}
\end{table}

Table~\ref{tab:ablation} reports the performance of various instance sampling strategies.
We evaluate two alternative sampling strategies: \textbf{CB Selection}, which replaces intra-cluster random sampling with centroid-based selection, and \textbf{S Selection}, which retains only one centroid representative per cluster. Both strategies reduce semantic diversity and lead to clear performance drops. CB Selection reaches only 51.62\% (CH) and 57.51\% (EN) R@1 on MimoTable, falling 5.18 and 2.62 points below \textsc{CGPT}. S Selection performs well on Chinese but drops to 57.03\% on English, showing limited cross-lingual robustness. In contrast, \textsc{CGPT}'s random intra-cluster sampling consistently yields strong results across languages, emphasizing the need to preserve semantic variation for reliable instance selection.

\section{Conclusion}
We introduced \textsc{CGPT}, a training framework that addresses the limited semantic coverage of zero-shot embedding models in table retrieval. By constructing semantically informed partial tables via K-means clustering and applying contrastive fine-tuning with hard negative sampling, \textsc{CGPT} substantially improves table representations. Across multiple benchmarks, it consistently surpasses QGpT, with an average R@1 gain of 16.54\%.  

\begin{acks}
This research was supported (in part) by NSTC 114-2634-F-005-002 - project Smart Sustainable New Agriculture Research Center (SMARTer).
\end{acks}

\bibliographystyle{ACM-Reference-Format}
\bibliography{reference}

\appendix
\section{Prompts}
\label{sec:prompts}
This appendix provides the exact prompt template used for synthetic query generation in Section~3.2. 

\begin{tcolorbox}[
  colback=gray!5,
  colframe=gray!75!black,
  title=Prompt,
  breakable,
  enhanced,
  verbatim,
  left=3mm,
  right=3mm,
  top=2mm,
  bottom=2mm
]
You are given a table chunk with the following content:
{table\_chunk}

Your Task:
Generate {questions\_per\_chunk} diverse questions that would retrieve this specific table chunk.
The questions should be based on the actual content shown in the table above.

Question Types to Cover:

1. Entity-specific query

2. Temporal query

3. Comparison/Ranking query

4. Aggregation query

5. Complex reasoning query

Important Requirements:

- Use natural, conversational language

- Make questions specific to the actual content shown in the table

- Reference real values from the table when possible

- Questions should be answerable by looking at this table chunk

- Language: {lang}

Output Format (JSON only):

{

  "questions": ["question1", "question2", "question3", ...]
  
}

Generate {questions\_per\_chunk} questions now:
\end{tcolorbox}

\end{document}